\documentclass[aps,prl,twocolumn,superscriptaddress,floatfix,showpacs]{revtex4}
\usepackage{graphicx}
\usepackage{dcolumn}
\usepackage{amsmath}
\usepackage{color}

\begin{document}
\title{Calculated Phonon Spectra of Paramagnetic Iron at the $\alpha$--$\gamma$ Phase Transition }

\author{I. Leonov}
\affiliation{Theoretical Physics III, Center for Electronic Correlations and Magnetism,
Institute of Physics, University of Augsburg, 86135 Augsburg, Germany}
\author{A. I. Poteryaev}
\affiliation{Institute of Metal Physics, Sofia Kovalevskaya Street 18, 620219 Yekaterinburg
GSP-170, Russia}
\affiliation{Institute of Quantum Materials Science, 620107 Yekaterinburg, Russia}
\author{V. I. Anisimov}
\affiliation{Institute of Metal Physics, Sofia Kovalevskaya Street 18, 620219 Yekaterinburg
GSP-170, Russia}
\author{D. Vollhardt}
\affiliation{Theoretical Physics III, Center for Electronic Correlations and Magnetism,
Institute of Physics, University of Augsburg, 86135 Augsburg, Germany}

\begin{abstract}
We compute lattice dynamical properties of iron at the bcc-fcc phase transition using dynamical 
mean-field theory implemented with the frozen-phonon method. Electronic correlations are found to have a 
strong effect on the lattice stability of paramagnetic iron in the bcc phase. Our results for the 
structural phase stability and lattice dynamical properties of iron are in good agreement with experiment.
\end{abstract}

 \maketitle

Elemental iron (Fe) is an exceptionally important metal for modern-day industry.
For example, iron is the main ingredient of steel. Therefore a detailed understanding of
the properties of iron is essential for many technological applications.
The remarkable macroscopic properties of iron are the result of a complex interplay between electronic and lattice degrees of 
freedom on the microscopic level. As a consequence iron exhibits a rich phase diagram with at least four allotropic forms. 
At ambient pressure the ground state is ferromagnetic and has a bcc crystal structure ($\alpha$ phase). Upon heating iron 
becomes paramagnetic at $T_C \sim 1043$ K, while retaining its bcc structure. Only when the temperature is further increased 
to $T_{\rm struct} \sim 1185$ K does the lattice transform to a fcc structure ($\gamma$ phase) \cite{BH55}. Under pressure $\alpha$
iron makes a transition to a paramagnetic hcp structure ($\epsilon$ phase) at $\sim 11$ GPa.
Even today, and in spite of long-term intensive research, the phase diagram of iron at high temperatures and pressures, as well 
as several key properties, are still poorly understood.


State-of-the-art methods for the calculation of the electronic structure provide a qualitatively
correct description of the equilibrium crystal structure and the lattice dynamical properties of the
ferromagnetic $\alpha$ phase of iron. Various other properties of iron can also be 
understood on the basis of band structure calculations \cite{SP91}.
However, applications of these techniques to describe, e.g., the $\alpha$-$\gamma$ phase transition
in iron, do not lead to satisfactory results. This is mainly due to the presence of local moments above 
$T_C$ which cannot be treated adequately by conventional band structure techniques.
Namely, these methods predict a simultaneous transition of the structure and the 
magnetic state at the bcc-fcc phase transition while, in fact, the bcc-to-fcc phase transition occurs 
only $\sim 150$ K above $T_C$. In addition, the elastic and dynamical stability of the bcc phase is 
found to depend sensitively on the value of the magnetization. For example, in the absence of a magnetization 
conventional methods find bcc iron to be mechanically unstable \cite{HC02}. Clearly, an overall explanation 
of the properties of iron requires a formalism which can take into account the existence of local 
moments above $T_C$.


The LDA/GGA+DMFT approach, a combination of the \emph{ab initio} local
density approximation (LDA) or generalized gradient approximation (GGA)
of the density functional theory with dynamical mean-field theory (DMFT),
allows one to determine the electronic and structural properties of materials
with correlated electrons in both their paramagnetic and magnetically ordered
states \cite{DMFT,LDA+DMFT}.
Applications of LDA+DMFT have shown to provide a good quantitative
description of the magnetization and the susceptibility of $\alpha$ iron as a
function of the reduced temperature $T/T_C$ \cite{DMFTcalc+,LK01}.
It was found that the formation of local moments in the paramagnetic $\alpha$ phase
is governed by the $e_g$ electrons which is accompanied by non-Fermi liquid
behavior \cite{KPE10}. This supports the results obtained with the $s$-$d$
model for the $\alpha$ phase of iron \cite{IKT93}.
The LDA/GGA+DMFT approach implemented with plane-wave pseudopotentials \cite{LB08,TL08}
has been recently employed to compute the equilibrium crystal structure and phase
stability of iron at the $\alpha$-$\gamma$ phase transition \cite{LP11}.
The bcc-to-fcc phase transition was found to take place at $\sim 1.3~T_C$, i.e.,
well above the magnetic transition, in agreement with experiment.
However, the lattice dynamical properties of \emph{paramagnetic} iron at the
$\alpha$-$\gamma$ phase transition, a problem posing a great theoretical and
experimental challenge, remained unexplored.


In this Letter, we determine the structural phase stability and lattice dynamics
of paramagnetic iron at finite temperatures by employing the LDA/GGA+DMFT
computational scheme. The approach is implemented with plane-wave
pseudopotentials which allows us to compute lattice transformation effects
caused by electronic correlations \cite{LB08,TL08,LP11}.
We employ this \emph{ab initio} computational scheme in combination with
the method of frozen phonons \cite{frzn_phns} to calculate the temperature dependent
phonon dispersion relations and phonon spectra of
paramagnetic iron at the bcc-fcc phase transition, a computation which
was not possible up to now.

\begin{figure}[tbp!]
\centerline{\includegraphics[width=0.5\textwidth,clip]{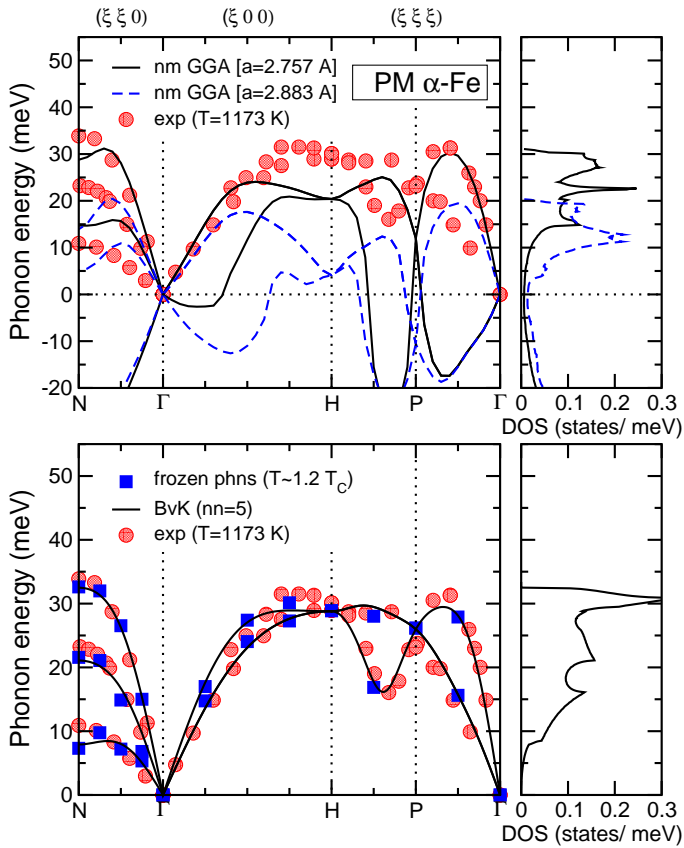}}
\caption{(color online)
Phonon dispersion curves and corresponding phonon density of states
of paramagnetic bcc Fe as calculated within the non-magnetic GGA (top) and 
DMFT (bottom). The results are compared with neutron
inelastic scattering measurements at 1173 K \cite{NP97}.
}
\label{fig:phns_bcc}
\end{figure}

\begin{figure}[tbp!]
\centerline{\includegraphics[width=0.5\textwidth,clip]{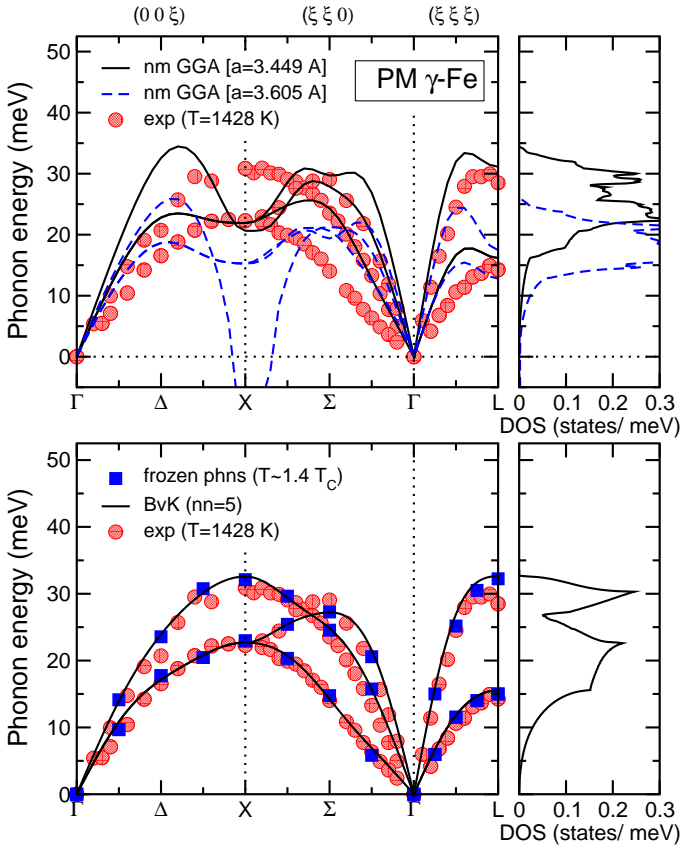}}
\caption{(color online)
Phonon dispersion curves and corresponding phonon density of states
of paramagnetic fcc Fe as calculated within the non-magnetic GGA (top) and 
DMFT (bottom). The results are compared
with neutron inelastic scattering measurements at 1428 K \cite{ZS87}.
}
\label{fig:phns_fcc}
\end{figure}

We first compute the electronic structure of iron within the non-magnetic GGA
by employing the plane-wave pseudopotential approach \cite{PSEUDO} and 
determine the equilibrium lattice constant $a$ for the bcc and fcc structures.
Overall, our results agree well with previous band structure
calculations \cite{SP91}, e.g., we find $a = 2.757$ \AA\ for the bcc phase and
$a = 3.449$ \AA\ for the fcc phase of iron. It should be noted that these
values are considerably smaller than those observed in the experiment \cite{BH55}.
%
Using the equilibrium lattice constants calculated above, we compute lattice
dynamical properties of iron, by using the first-principles linear response 
method \cite{PSEUDO}. 
In Fig.~\ref{fig:phns_bcc} (top) we show the results for the phonon
dispersion curves and the corresponding phonon density of states of bcc Fe.
The non-magnetic GGA finds the bcc lattice to be dynamically unstable with 
negative elastic constants C$_{11}$ and C$^\prime$ (see Table~\ref{table1}),
in accordance with previous calculations \cite{HC02}. 
By contrast, the same method finds the fcc lattice structure (with $a = 3.449$ \AA) 
to be mechanically stable (see Fig.~\ref{fig:phns_fcc}) and the calculated 
phonon frequencies to deviate significantly from the experimental data. 
Indeed, the non-magnetic GGA finds a strong softening 
of the longitudinal [00$\xi$] mode at the $X$ point by $\sim 30$\%.
Furthermore, all calculated elastic constants are 2-3 times larger than in the 
experiment (see Table~\ref{table1}).
%
%
Overall, calculations within the non-magnetic GGA find lattice dynamical properties
which are in disagreement with experiment.
Apparently, conventional band structure techniques cannot explain the experimentally 
observed structural phase stability of paramagnetic iron at the bcc-fcc phase transition,
since they do not describe electronic correlations adequately.


To include the effect of electronic correlations, we compute an effective
low-energy Hamiltonian for the partially filled Fe $sd$ orbitals \cite{TL08,AK05}.
We employ results of band structure calculations for iron performed with the 
non-magnetic GGA to construct a basis of atomic-centered symmetry-constrained Wannier functions
for the Fe $sd$ orbitals \cite{TL08,AK05}. The corresponding first-principles
multiband Hubbard Hamiltonian has the form
\begin{eqnarray}
{\hat H} & = & {\hat H_\mathrm{GGA}} + \frac{1}{2}\sum_{imm',\sigma\sigma'}
U^{\sigma \sigma'}_{mm'} \hat n_{im\sigma} \hat n_{im'\sigma'} - {\hat H_\mathrm{DC}}
\label{eqn:hamilt}
\end{eqnarray}
where $\hat n_{im\sigma} = \hat c^\dagger_{im\sigma} \hat c_{im\sigma}$,
and $\hat c^\dagger_{im\sigma}$ ($\hat c_{im\sigma}$) creates (destroys)
an electron with spin $\sigma$ in a Wannier orbital $m$ at site $i$.
Here ${\hat H_\mathrm{GGA}}$ is the effective low-energy Hamiltonian
in the basis of Fe $sd$ Wannier orbitals, and
${\hat H_\mathrm{DC}}$ is a double-counting correction which accounts for
the electronic correlations already described by the GGA.
We use $U=1.8$ eV and $J=0.9$ eV in our calculations as obtained by
previous theoretical and experimental estimations \cite{U_Fe}. 
To solve the many-body Hamiltonian (1) we employ the DMFT together 
with quantum Monte Carlo (QMC) simulations with the Hirsch-Fye
algorithm \cite{HF86}.


We now compute the lattice dynamics of paramagnetic iron using the GGA+DMFT 
approach \cite{DMFT,LDA+DMFT} in combination with the method of frozen
phonons \cite{frzn_phns}.
The phonon frequencies are calculated by introducing
a small set of displacements in the corresponding supercells of the equilibrium
lattice which results in a total energy difference with respect to the undistorted structure.
To this end we first perform a direct structural optimization and compute the
equilibrium lattice constant of iron. We focus on the lattice dynamical properties
of iron near the bcc-to-fcc phase transition. Namely, we perform our calculations
at temperatures $T=1.2~T_C$ and $1.4~T_C$, which are below and
above the temperature $T_{\rm struct} \sim 1.3~T_C$ where the structural phase 
transition occurs \cite{LP11}.
We calculate the total energy of the paramagnetic bcc and fcc structures as
functions of the volume and thereby determine the equilibrium lattice constants 
for the temperatures mentioned above (see Table~\ref{table1}).
Our results for the equilibrium lattice constants, which now include the effect of 
electronic correlations, agree well with experiment:
for the bcc phase we find $a = 2.883$ \AA, which is only $< 1$ \% smaller than
the experimental value \cite{BH55}, and for the fcc phase $a = 3.605$ \AA, which is
only $< 2$ \% smaller than in experiment \cite{BH55}. In the following we calculate
the lattice dynamics of paramagnetic bcc and fcc iron, respectively.


\underline{Paramagnetic bcc iron:} At $T \sim 1.2~T_C$ we find the bcc phase to be 
energetically favorable, i.e., thermodynamically stable, with a difference 
in the total energy between the bcc and fcc phases of
$\Delta E \equiv E_{\rm fcc}-E_{\rm bcc}\sim 25$ meV/at.
In Fig.~\ref{fig:phns_bcc} (bottom)
we present our results for the phonon dispersion relations and phonon spectra.
We note that
these calculations are performed for the equilibrium volume (with $a = 2.883$ \AA)
computed at this particular temperature.
%
%
To evaluate the phonon frequencies at arbitrary wave vector of the Brillouin
zone we performed lattice dynamical calculations on the basis of a Born-von K\'arm\'an
model with interactions expanded up to the 5-th nearest-neighbor shell.
The calculated phonon dispersions of the bcc phase of iron show the typical behavior
of a bcc metal with an effective Debye temperature $\Theta_{\rm D} \sim 458$ K.
The phonon frequencies are overall positive
which implies mechanical stability of the bcc lattice structure at $T \sim 1.2~T_C$,
i.e., well above the Curie temperature. This result is quite different from
that obtained with the non-magnetic GGA which finds the bcc lattice to be dynamically unstable
(even for the equilibrium lattice constant $a = 2.883$ \AA\ which almost
coincides with experiment). Therefore this approximation cannot explain the bcc-to-fcc phase transition
in paramagnetic iron.
Most importantly, our results clearly demonstrate the crucial importance of electronic
correlations to explain both the \emph{thermodynamic} and the \emph{lattice dynamical}
stability of the paramagnetic bcc phase of iron.

Overall, the structural phase stability, equilibrium lattice constant,
and phonon frequencies of bcc iron obtained by GGA+DMFT are in remarkably
good agreement with the experimental data which were taken at nearly the same reduced
temperature $T/T_C$ \cite{BH55,NP97}.
Nevertheless, we notice a weak anomaly in the transverse $T_1$ acoustic mode along 
the [$\xi \xi$0] direction, indicating that at $T \sim 1.2~T_C$ the bcc phase may be close
to an instability. This result can be ascribed to a dynamical precursor effect of the
bcc-to-fcc phase transition.
We note that a similar behavior of the $T_1$ [$\xi\xi$0] phonon mode was
found to occur in the $\delta$ iron at high temperatures \cite{LJ10}.
It appears that the temperature driven bcc-to-fcc phase transition in paramagnetic 
iron differs from the pressure driven bcc-to-hcp phase
transition, where neutron studies found no dynamical precursor effects \cite{KB00}.


In addition, using our result for the phonon dispersions, we compute elastic
properties of paramagnetic iron. The elastic constants C$_{11}$, C$_{12}$,
and C$_{44}$ (due to the cubic symmetry there are only three independent parameters) 
are obtained from the estimates of the corresponding sound velocities
along the [$\xi$00] and [$\xi\xi$0] directions. Our results for the elastic constants are
summarized in Table~\ref{table1} where they are compared with results from non-magnetic
GGA and experimental data.
The elastic constants obtained by GGA+DMFT are seen to agree well with the available
experiments.
%
\begin{table}[tbp!]
\caption{ Comparison between calculated and experimental elastic constants
(in 10$^{12}$ dyn/cm$^2$) of iron. The calculated equilibrium and experimental
lattice constants are presented in the last column.}
\begin{ruledtabular}
\begin{tabular}{lccccccc}
Method & Phase & $T/T_C$ & C$_{11}$ & C$_{44}$ & C$_{12}$ & C$^\prime$ & a, \AA\ \\
\hline
nm GGA  & bcc & $-$  & -0.15 & 1.19 & 4.92 & -1.87 & 2.757  \\
DMFT    & bcc & 1.2  &  2.30 & 1.27 & 1.57 &  0.36 & 2.883  \\
Exp.\cite{BH55,bcc_elstc_const} & bcc & 1.1  &  1.92 & 1.24 & 1.71 & 0.10 & 2.897 \\
\hline
nm GGA  & fcc & $-$  & 3.21  & 1.97 & 1.74 &  0.73 & 3.449  \\
DMFT    & fcc & 1.4  & 2.10  & 1.38 & 1.61 &  0.25 & 3.605  \\
Exp.\cite{BH55,ZS87} & fcc & 1.4  & 1.54 & 0.77 & 1.22 & 0.16 & 3.662 \\
\end{tabular}
\end{ruledtabular}
\label{table1}
\end{table}


\underline{Paramagnetic fcc iron:} Next we calculate the lattice dynamical properties of the
paramagnetic fcc phase of iron. Our calculations of the structural phase stability
find that upon heating the fcc phase is energetically favorable for $T>T_{\rm struct}$.
The total energy difference between the bcc and fcc phase is $\Delta E \sim -20$ meV/at
at $T \sim 1.4~T_C$.
To prove the mechanical stability of the fcc phase at this temperature we
now compute the lattice dynamics of the fcc phase of iron.
Our results for the phonon dispersion relations and phonon spectra, which were
obtained for the equilibrium lattice constant $a = 3.605$ \AA\, are shown in
Fig.~\ref{fig:phns_fcc} (bottom).
The effective Debye temperature at $T \sim 1.4~T_C$ is found
to be $\Theta_{\rm D} \sim 349$ K.
%
%
The phonon frequencies are overall positive,
implying mechanical stability of the fcc lattice structure at $T \sim 1.4~T_C$.
This qualitatively agrees with the results of non-magnetic GGA calculations which
(for the GGA equilibrium volume) predict the fcc lattice structure to be mechanically
stable. However, the GGA energy for fcc iron is higher than that
for the close-packed hcp structure.
By contrast, the GGA+DMFT calculations find the simultaneous \emph{thermodynamic} and
\emph{lattice dynamical} stability of the paramagnetic fcc phase of iron at
$T \sim 1.4~T_C$, in accordance with experiment.
Our results for the structural phase stability, equilibrium lattice constant, 
and phonon frequencies agree remarkably well with the available experimental
data taken at nearly the same reduced temperature $T \sim 1.4~T_C$ \cite{BH55,ZS87}.
It is important to note that the application of the non-magnetic GGA to fcc iron finds 
phonon frequencies which differ considerably from experiment.
These findings clearly demonstrate the importance of electronic
correlations for the lattice dynamical properties of fcc iron.


In conclusion, we employed the GGA+DMFT computational scheme 
to determine the equilibrium crystal structure and lattice dynamics
of iron at the bcc-fcc phase transition.
The calculated structural phase stability and lattice dynamical
properties of iron near the bcc-to-fcc phase transition are in overall good
\emph{quantitative} agreement with experiment.
Most importantly, our calculations explain both the
\emph{thermodynamic} and the \emph{lattice dynamical} stability of the paramagnetic
bcc phase of iron below the bcc-fcc structural phase transition.
In particular, electronic correlations are found to be crucial for an 
explanation of the equilibrium crystal structure and lattice dynamics of iron.
%

\begin{acknowledgments}
We thank Yu. N. Gornostyrev, F. Lechermann, and A. I. Lichtenstein for valuable
discussions. Support by the Deutsche Forschergemeinschaft through TRR 80 (I.L.) 
and FOR 1346 (V.I.A., D.V.), as well as by RFFI-10-02- 00046a, and
NSH 4711.2010.2 is gratefully acknowledged.

\end{acknowledgments}


\begin{thebibliography}{99}

\bibitem{BH55} Z. S. Basinski, W. Hume-Rothery, and A. L. Sutton, Proc. R. Soc. London, Ser. A \textbf{229}, 459 (1955).
R. W. G. Wyckoff, Crystal structure, vol. 1 (Wiley, New York, 1963).

\bibitem{SP91} D. J. Singh, W. E. Pickett, and H. Krakauer, Phys. Rev. B \textbf{43}, 11628 (1991);
C. Amador, W. R. L. Lambrecht, and B. Segall, \emph{ibid.} \textbf{46}, 1870 (1992);
L. Stixrude, R. E. Cohen, and D. J. Singh, \emph{ibid.} \textbf{50}, 6442 (1994);
A. Dal Corso and S. de Gironcoli, \emph{ibid.} \textbf{62}, 273 (2000);
S. V. Okatov \emph{et al.}, Phys. Rev. B \textbf{79}, 094111 (2009).

\bibitem{HC02} H. C. Hsueh \emph{et al.}, Phys. Rev. B \textbf{66}, 052420 (2002).

\bibitem{DMFT} W. Metzner and D. Vollhardt, Phys. Rev. Lett. \textbf{62}, 324 (1989);
G. Kotliar and D. Vollhardt, Phys. Today \textbf{57}, No. 3, 53 (2004);
A. Georges \emph{et al.}, Rev. Mod. Phys. \textbf{68}, 13 (1996).

\bibitem{LDA+DMFT}
V. I. Anisimov \emph{et al.}, J. Phys.: Cond. Matt. \textbf{9}, 7359 (1997);
A. I. Lichtenstein and M. I. Katsnelson, Phys. Rev. B \textbf{57}, 6884 (1998);
K. Held \emph{et al.}, 
Phys. Status Solidi B \textbf{243}, 2599 (2006); K. Held, Adv. Phys. \textbf{56}, 829 (2007);
G. Kotliar \emph{et al.}, Rev. Mod. Phys. \textbf{78}, 865 (2006).

\bibitem{DMFTcalc+} M. I. Katsnelson and A. I. Lichtenstein Phys. Rev. B \textbf{61}, 8906 (2000);
L. Chioncel \emph{et al.}, Phys. Rev. B \textbf{67}, 235106 (2003);
J. Braun \emph{et al.}, Phys. Rev. Lett. \textbf{97}, 227601 (2006);
A. Grechnev \emph{et al.}, Phys. Rev. B \textbf{76}, 035107 (2007);
S. Chadov \emph{et al.}, Europhys. Lett. \textbf{82}, 37001 (2008).

\bibitem{LK01}
A. I. Lichtenstein, M. I. Katsnelson, and G. Kotliar, Phys. Rev. Lett. \textbf{87}, 067205 (2001).

\bibitem{KPE10} A. A. Katanin \emph{et al.}, Phys. Rev. B \textbf{81}, 045117 (2010).

\bibitem{IKT93} J. B. Goodenough, Phys. Rev. \textbf{120}, 67 (1960);
V. Yu. Irkhin, M. I. Katsnelson, and A. V. Trefilov, J. Phys.: Cond. Matt. \textbf{5}, 8763 (1993).

\bibitem{LB08} I. Leonov \emph{et al.}, Phys. Rev. Lett. \textbf{101}, 096405 (2008);
I. Leonov \emph{et al.}, Phys. Rev. B, \textbf{81}, 075109 (2010).

\bibitem{TL08} G. Trimarchi \emph{et al.}, J. Phys.: Condens. Matter \textbf{20}, 135227 (2008);
Dm. Korotin \emph{et al.}, Eur. Phys. J. B \textbf{65}, 91 (2008).

\bibitem{LP11} I. Leonov \emph{et al.}, Phys. Rev. Lett. \textbf{106}, 010645 (2011).

\bibitem{frzn_phns} H. T. Stokes, D. M. Hatch, and B. J. Campbell, (2007). ISOTROPY, stokes.byu.edu/isotropy.html.

\bibitem{PSEUDO} S. Baroni \emph{et al.}, URL http://www.pwscf.org;
S. Baroni \emph{et al.}, Rev. Mod. Phys. \textbf{73}, 515 (2001);
P. Giannozzi \emph{et al.}, J. Phys.: Condens. Matter \textbf{21},
395502 (2009).

\bibitem{NP97} J. Neuhaus, W. Petry, and A. Krimmel, Physica B \textbf{234-236}, 897 (1997).

\bibitem{bcc_elstc_const} These elastic constants were extracted using a Born-von K\'arm\'an
model with interactions expanded up to the 5-th nearest-neighbor shell for data taken from
Ref.~\cite{NP97}.

\bibitem{ZS87} J. Zarestky and C. Stassis, Phys. Rev. B \textbf{35}, 4500 (1987).


\bibitem{AK05} V. I. Anisimov \emph{et al.}, Phys. Rev. B \textbf{71}, 125119 (2005).

\bibitem{U_Fe} E. Antonides, E. C. Janse, and G. A. Sawatzky, Phys. Rev. B \textbf{15}, 1669 (1977);
G. Tr\' eglia \emph{et al.}, J. Phys. C \textbf{14}, 4347 (1981);
M. M. Steiner, R. C. Albers, and L. J. Sham, Phys. Rev. B \textbf{45}, 13272 (1992);
M. Cococcioni and S. de Gironcoli, Phys. Rev. B \textbf{71}, 035105 (2005);
I. Yang, S. Y. Savrasov, and G. Kotliar, Phys. Rev. Lett. \textbf{87}, 216405 (2001);
N. L. Stoji\' c and N. L. Binggeli, J. Magn. Magn. Matter. \textbf{320}, 100 (2008).

\bibitem{HF86} J. E. Hirsch and R. M. Fye, Phys. Rev. Lett \textbf{56}, 2521
(1986).

\bibitem{KB00} S. Klotz and M. Braden, Phys. Rev. Lett \textbf{85}, 3209 (2000).

\bibitem{LJ10} W. Luo \emph{et al.}, PNAS \textbf{107}, 9962 (2010).

\end{thebibliography}
\end{document}